\documentclass[12pt, preprint]{aastex}
\def\pppm{\rm P^3M}
\def\mpchi{\,h^{-1}{\rm {Mpc}}}
\def\mpchii{\,h{\rm {Mpc}^{-1}}}

\def\msun{\,h^{-1}{\rm M_\odot}}

\def\etl{et al.}
\begin{document}

\title{The dependence of dark halo clustering on the formation epoch
and the concentration parameter}
\author{Y. P. Jing}
\affil{Shanghai Astronomical Observatory, Nandan Road 80, Shanghai, 200030,
China}
\author{Yasushi Suto}
\affil{Department of Physics and Research Center for the Early
Universe (RESCEU)\\ School of Science, University of Tokyo, Tokyo
113-0033, Japan}
\author{and}
\author{H.J. Mo}
\affil{Department of Astronomy, University of Massachusetts,
Amherst MA 01003-9305} 

\affil{e-mail: ypjing@shao.ac.cn,suto@phys.s.u-tokyo.ac.jp,
hjmo@nova.astro.umass.edu}

\begin{abstract}
 We examine the age-dependence of dark matter halo clustering in an
unprecedented accuracy using a set of 7 high-resolution cosmological
simulations each with $N=1024^3$ particles.  We measure the bias
parameters for halos over a large mass range using the
cross-power-spectrum method that can effectively suppress the random
noise even in the sparse sampling of the most massive halos. This
enables us to find, for the first time, that younger halos are more
strongly clustered than older ones for halo masses $M>10M_{\ast}$,
where $M_\ast$ is the characteristic nonlinear mass scale. For
$M<M_{\ast}$, our results confirm the previous finding of Gao et
al. that older halos are clustered more strongly than the younger
ones. We also study the halo bias as a function of halo concentration,
and find that the concentration dependence is weaker than the age
dependence for $M<M_{\ast}$, but stronger for $M\ga 50 M_{\ast}$.  The
accurate and robust measurement of the age dependences of halo bias
points to a limitation of the simple excursion set theory which
predicts that the formation and structure of a halo of given mass is
independent of its environment.
\end{abstract}


\section{Introduction}

The formation of dark matter halos plays a central role in the studies
of galaxy formation as well as of the large-scale structure of
the universe. A widely used analytical theory for the formation of 
halos is the extended Press-Schechter (PS thereafter) formalism
\citep{bond91,bower91}, which can be used to model the
mass function of dark halos \citep{ps74}, the bias parameter
as a function of halo mass \citep[][hereafter MW96]{mw96}, 
as well as the formation histories of dark halos \citep{LC1993,LC1994}. 
In the simplest excursion-set model of \citet{bond91}, 
the physical properties of dark halos are expected to depend 
only on halo mass, but not on large-scale environments. 
However, it has been known for almost a decade that neither the 
PS mass function nor the MW96 bias model matches well $N$-body 
results for halo masses $M < M_\ast$, where $M_\ast$ is defined
such that the rms linear density fluctuation within a sphere of mass
$M_\ast$ is $\delta_c=1.68$ \citep{jing98,lee98,sl99,porciani99,j991,st99}. 
\citet{smt01} proposed an ellipsoid collapse model to replace 
the spherical collapse model in the PS theory, and found a better 
agreement between the theory and $N$-body simulations 
in both the mass function and the bias function.
Earlier investigations with cosmological $N$-body simulations did 
not find any strong environmental dependence of halo properties, such as 
the spin parameter, concentration and formation time 
\citep{lemson99,percival03}. However, the poor resolutions and small 
dynamical ranges covered in these simulations make it difficult 
to detect any signals that are relatively weak or outside the 
dynamical range, and so a significant age-dependence of halo 
clustering cannot be ruled out. Note also that there were 
earlier attempts to develop empirical models for the age-dependence 
of halo clustering \citep{taruya00,Hamana01,YJTS01}.

 With the use of a very large $N$-body simulation, \citet{gao05} 
recently demonstrated convincingly that the clustering strength 
of dark halos does depend on halo formation time.   
In particular, they found that this dependence is strong 
only for halos with $M\ll M_\ast$ and becomes very weak for $M>M_\ast$.  
This also explains why \citet{percival03} could not detect 
an age dependence, because they focused on halos with $M>M_\ast$ 
at high redshifts. \citet{wang06} have examined the physical 
process that may be responsible for the age dependence of 
halo clustering. They found that halos embedded in 
dense environments accrete mass less efficiently 
than the spherical collapse model predicts, because the 
matter to be accreted is `heated' by the large-scale structure
\citep[like the pancake heating considered in, 
e.g.,][]{mo05,lee06}. This explains why the old population
of small halos has a higher bias than the young population 
of the same mass. It also qualitatively explains why the PS mass 
function and bias functions deviate from $N$-body simulations 
at small masses.

It is well known that the concentration parameter of halos at a given
mass, $c(M)$, has a broad log-normal distribution, and is
correlated with the halo formation epoch $z_{\rm form}$ in such a way
that halos of earlier formation have a higher concentration
\citep{jing00,bullock01,js02,Wechsler02,zhao03a,zhao03b}.  
Therefore the age dependence of halo bias should yield a 
concentration dependence of halo bias. Such dependence is
indeed observed in recent $N$-body simulations.     
\citet{Wechsler06} found that, for masses $M<M_\ast$,
halos with higher concentrations are more strongly clustered, 
as expected from the age dependence and the correlation 
between the age and concentration of dark halos.
More interestingly, \citet{Wechsler06} found reversed
concentration dependence for $M>M_\ast$, in the sense 
that halos with higher concentrations are actually less biased.
This result was not seen by \citet{gao05} who explored the age 
dependence only for $M\le 10M_\ast$. More recently, \citet{white06} 
examined both the age dependence and concentration dependence 
of halo clustering for massive halos. While they found 
concentration dependence similar to that found by 
\citet{Wechsler06}, they did not detect any significant
dependence on formation epoch. Note, however, that
\citet{Wechsler06} used halos identified at different 
redshifts to increase the dynamical range probed by their 
simulations, while both \citet{gao05} and \citet{white06}
used halos identified at the same time. It is unclear how 
to make a detailed comparison between the different results.
 
The environmental dependence of halo formation and structure
has important implications not only for improving the PS theory 
but also for improving semi-analytical models of galaxy 
formation based on PS merger trees and current   
halo occupation distribution (HOD) models. In this {\it Letter}, 
we use a large set of cosmological simulations, each with $1024^3$ 
particles, to investigate in detail the concentration and
formation epoch dependence of halo clustering. 

\section{Simulations}

 The model considered here is a canonical spatially-flat Cold 
Dark Matter (CDM) model with the density parameter $\Omega_m=0.268$, 
the cosmological constant $\Omega_\Lambda=0.732$, the Hubble constant 
$h=0.71$, and the baryon density parameter $\Omega_b=0.045$. 
The primordial density field is assumed to be Gaussian with 
a scale-invariant power spectrum $\propto k$. For the
linear spectrum, we adopt the fitting transfer function of
\citet{eh98}, and the normalization is set by $\sigma_8=0.85$, 
where $\sigma_8$ is the present linear RMS density
fluctuation within a sphere of radius $8\mpchi$. 

We use an upgraded version of the Particle-Particle-Particle-Mesh
($\pppm$) code of \citet{js98,js02} to simulate structure formation in
the universe. The code has now incorporated the multiple level $\pppm$
gravity solver for high density regions \citep{js00}. In order to have a
large mass resolution range, we run a total of 7 simulations with $1024^3$
particles in different simulation boxes (Table \ref{tab:simu}). The
simulations were run on an SGI Altix 350 with 16 processors 
with OPENMP parallelization in Shanghai Astronomical Observatory. 
We have 4 realizations with boxsize $1800\mpchi$ in order to reliably 
measure the bias of the most massive halos, which is the focus of 
the present {\it Letter}.

Dark matter halos are identified using the standard Friends-of-Friends
algorithm with a linking length $b$ equal to $0.2$ times the mean
particle separation.  Unbound particles (with positive binding energy)
are excluded.  We use halos containing 100 particles or more, and our
analysis covers halos with masses ranging from $2\times 10^{11}\msun$
to $10^{15}\msun$.  

\section{Clustering of dark matter halos}

\subsection{Cross power spectrum and the halo bias}

 Once the number density field of dark matter halos,
$n_h({\bf r})$, and the density field of dark matter 
$\rho_m({\bf r})$, are given, the conventional way to estimate 
the halo bias factor $b$ is to use the definition,  
$b\equiv (\xi_{hh}(r)/ \xi_{mm}(r))^{1/2}$, 
where $\xi_{hh}(r)$ is the two-point correlation function 
of dark halos, and $\xi_{mm}(r)$ that of dark matter.
Here we adopt a slightly different approach by  
using the cross power spectrum of \citet{j991}.
We first Fourier transform both $n_h({\bf r})$ and
$\rho_m({\bf r})$ into $n_h({\bf k})$ and $\rho_m({\bf k})$, 
and then measure the bias factor through
\begin{equation}
\label{eq:bcps}
b\equiv \frac{P_{hm}(k)}{P_{mm}(k)}=\frac{{\overline \rho_m}}{{\overline
n_h}}\frac{\langle n_h({\bf k})\rho_m({\bf
k})\rangle}{\langle \rho_m({\bf k})\rho_m({\bf k})\rangle}\,.
\end{equation}
As shown by \citet{j991}, this method suppresses 
Poisson noise due to a limited number of dark halos 
more effectively than the correlation method, because the 
number of dark particles is in general much larger than 
that of dark halos. This is particularly important 
for the present study, as we need to accurately determine the 
bias for massive halos whose number density is low.  
We have estimated the bias factors for 
the four realizations of the $1800 \mpchi$ box 
simulations, using both the correlation method and the cross 
power spectrum method. The average values of the bias over 
the four realizations obtained with the two methods are 
consistent, but the scatter with the correlation method
is about a factor of 2 larger than that with the cross power
spectrum. The cross power spectrum method is therefore preferred. 

 Errors on $b(k)$ are estimated following \citet{j991}: ten random
samples are generated for each simulated halo sample, 
and the scatter of $b(k)$ among them is used as the error in 
$b(k)$ for the real sample. We have compared the error so obtained
with that estimated from the scatter among the 4 realizations
of the $1800\mpchi$ box simulation. The errors given by the two 
methods are comparable.

\subsection{Dependence on the halo formation epoch}

\begin{figure}
\plotone{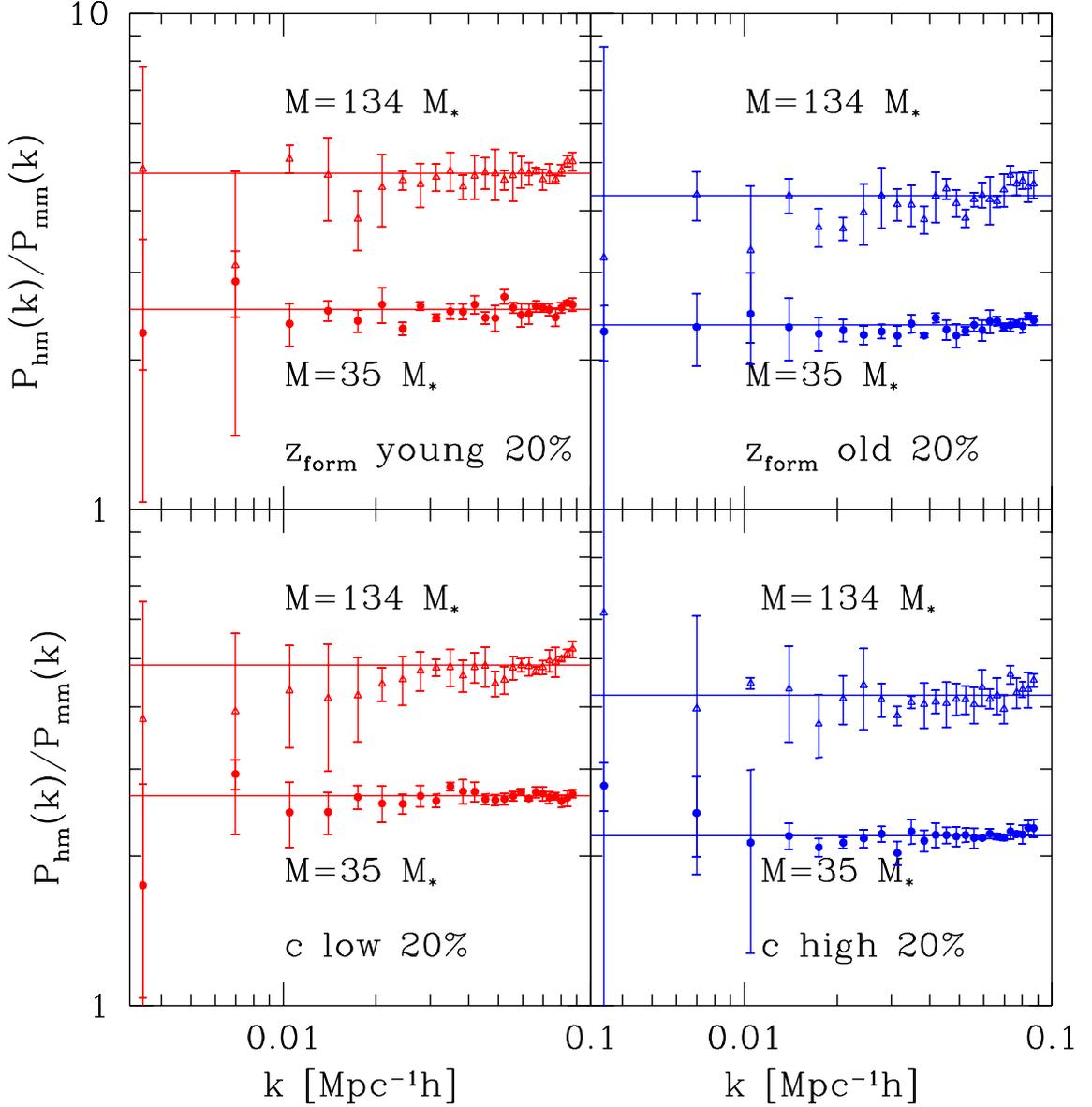} 
\caption{The ratio of the cross power spectrum $P_{hm}(k)$ of halos
with background dark matter to the matter power spectrum $P_{mm}(k)$
versus the wavelength in the linear regime. The top left and right
panels are for the 20\% youngest and 20\% oldest halos respectively,
where the age is defined as the formation redshift $z_f$. The bottom
left and right panels show the results for the 20\% halos that have
the highest and lowest concentrations respectively.  The mass of the
halos, in unit of the characteristic mass, is given in the panels.}
\label{fig:fig1}
\end{figure}

Our analysis focuses on the halo populations at the
present time, $z=0$. Halos are divided into mass bins
(specified by $[m_1,m_2]$ with $m_2=2m_1$) 
according to their masses. For each mass bin, we sort the halos 
into 5 populations according to formation redshift, $z_f$,  
which is defined as the redshift at which the mass of 
the most massive progenitor of a halo is equal to half of 
its mass. Although analysis was carried out for all the five 
populations, here we present the results only for the 20\% 
youngest halos with the smallest $z_f$ and for the 20\% oldest 
ones with the largest $z_f$. In what follows, these 
two populations are simply referred to as the young and 
the old populations, respectively. To ensure that 
the bias obtained here is in the linear regime, 
we only consider Fourier modes with $k<k_{max}$,
where $k_{max}=0.09\mpchii$ is a wavenumber at which     
the variance $\Delta^2(k_{max})
\equiv k_{max}^3P_{mm}(k_{max})/2\pi^2=0.25$. 
Figure \ref{fig:fig1} plots examples of the bias factor for 
massive halos with $M=35 M_{\ast}$ and $M=134 M_{\ast}$,
where $M_\ast =5\times10^{12}h^{-1}M_\odot$,
in the $1800\mpchi$ box simulations. The top-left panel shows 
the results for the young population, while the top-right 
panel shows the same measurement for the old population.
The results clearly show that the bias factors are 
nearly scale independent in the linear regime.  
The solid line in each panel is the mean value obtained by a 
least square fit to the data points at $k<k_{max}$. 
The bias factor for the young population is about 10\% higher than 
that for old population.

\begin{figure}
\plotone{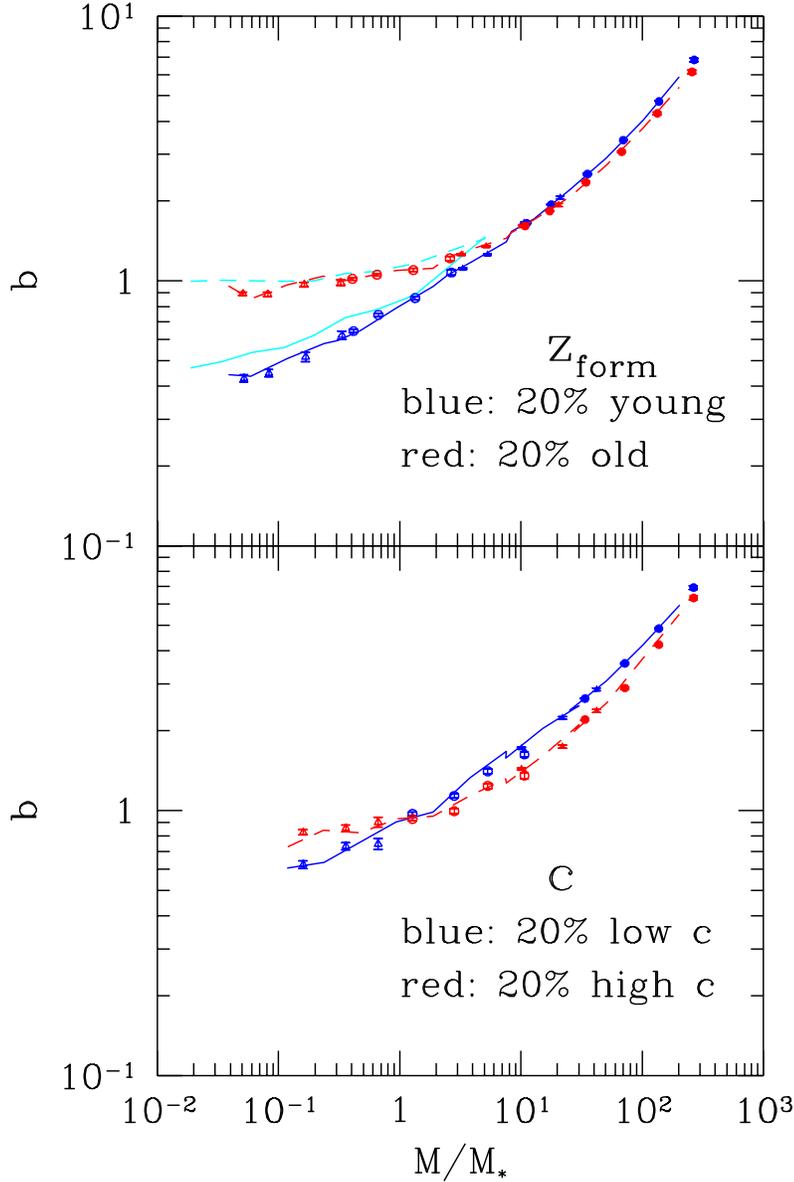} 
\caption{The bias factor of dark matter halos as a function of the
halo mass. In the upper panel, we show the dependence on the halo
formation epoch, with the symbols in red being for the 20\% oldest
halos and those in blue being for 20\% youngest halos. The cyan lines
are the results of \citet{gao05} for comparison. The lower panel shows
the dependence on the halo concentration, with the red and blue colors
being for those with the $20\%$ highest and lowest concentrations
respectively. The open triangles, open circles, filled triangles, and
filled circles are from the simulations of boxsize $300\mpchi$,
$600\mpchi$, $1200\mpchi$ and $1800\mpchi$ respectively. The solid and
dashed lines are not the lines connecting the data points, but are for
the results of young (or less concentrated) and old (or more
concentrated) halos respectively estimated for a very narrow mass bin
$M\pm 0.1M$.  }
\label{fig:fig2}
\end{figure}

We present the mean bias factor $b(M)$ as a function of mass 
obtained from different simulations in
the top panel of Figure \ref{fig:fig2}. For a given halo mass, 
the measurement of $b(M)$ is more accurate in a simulation of a
larger volume. Therefore we plot the data points for each simulation
only up to a halo mass which corresponds to 100 particles 
in the next larger simulation, except for the $1800 \mpchi$ box 
simulation, where we plot all the data points available. 
Note that the bias factors obtained from different simulations 
agree extremely well with each other, and so the set of simulations
used here allows us to explore the age dependence of halo
bias over four orders of magnitude in halo mass. 
For $M<3M_{\ast}$, the old population (red symbols) is more 
strongly biased (has stronger clustering) 
than the young population (blue symbols), which is in a
good quantitative agreement with the results obtained by 
\citet[][cyan lines]{gao05}. With the Millennium Simulation with a
boxsize of $500\mpchi$, \citet{gao05} was not able to explore 
accurately the age dependence of halo clustering for higher 
halo mass. Our results clearly show that the difference 
between the old and young populations decreases with
increasing halo mass up to $M\approx 10 M_{\ast}$, 
at which the young population starts to surpass the old 
population in the bias factor. For $M>20M_{\ast}$, 
the bias factor for the young population is about $10\%$ higher
than that of the old population, with weak dependence on halo mass. 
Although the difference in the bias factor between the old and 
young populations is small at the high mass end,
it is detected at a high statistical confidence ($\sim
10\sigma$).

\subsection{Dependence on halo concentration}
The concentration of each halo is obtained following the
fitting method of \citet{jing00}. The density distribution 
within each halo is fitted with a NFW profile to obtain the scale 
radius $r_s$, and the concentration is defined as $c=r_v/r_s$, 
where $r_v$ is the virial radius within which the mean density 
is 361 times the mean density of the universe.  
Here only halos with 320 particles or more are used, 
because the concentration may not be measured accurately
for halos containing smaller number of particles 
\citep[e.g.][]{Wechsler06}. 

  The lower two panels of Figure \ref{fig:fig1} show
the bias factor as a function of wavenumber   
for massive halos with $M=35 M_{\ast}$ and $M=134 M_{\ast}$. 
Results for the least concentrated 20\% are plotted
in the lower left panel, while those for the most concentrated 
20\% are in the lower right panel. Here again, 
the bias factor is almost scale-independent. The amplitude of 
the bias factor for massive halos clearly depends on concentration, 
with halos with higher concentration less strongly biased. 
To see how the concentration-dependence changes with halo mass, 
we plot in the lower panel of Figure \ref{fig:fig2} 
the bias factor for the most concentraed 20\% and the least 
concentrated 20\% of the halos in each of the mass bins.
Note that there is good agreement between simulations 
with different box sizes, suggesting that 320 particles
may be sufficient to sample the concentration for the
purpose of the present paper.Our results show clearly 
that the more concentrated halos have a larger bias for
$M<M_{\ast}$, but the trend is reversed for $M>M_{\ast}$, 
in qualitative agreement with the results in \citet{Wechsler06}. 

Comparing the results here with the dependence on the formation 
epoch (the upper panel), we see that the concentration
dependence is weaker than the age dependence for $M<M_{\ast}$, 
but stronger for $M \gg M_{\ast}$. The difference in the bias 
factor between the most concentrated 20\% and the least
concentrated 20\% is about 25\% for $M=10M_{\ast}$ - $100 M_{\ast}$,
larger than the 10\% between the youngest 20\% and oldest 20\%.
This may be why concentration dependence was but 
age dependence was not found for halos with $M>M_{\ast}$ in
previous investigations \citep{gao05,zhu06,Wechsler06,white06}.
Our results also show that the concentration dependence 
reverses almost exactly at $M=M_{\ast}$, while the reversal 
of age dependence occurs at a much larger mass, 
$M \approx 10 M_{\ast}$. Finally, the concentration 
dependence does not seem to become stronger with increasing mass
for $M>M_{\ast}$. In fact, the difference in the
bias factor between the most concentrated 20\% 
and the least concentrated 20\% is only about 10\% at
$250M_{\ast}$.

\subsection{The effect of finite mass bins}
\label{sect:massbin}

So far we have used a finite mass bin $[M_1,M_2]$ with $M_2=2M_1$
to study the age and concentration dependence of the bias 
for halos of a given mass. Since more massive halos on average have 
younger ages and smaller concentration, and since the bias factor 
increases strongly with halo mass at the high mass end, 
the use of a finite mass bin to represent a givn mass  
may artificially introduce age and concentration dependence.
To check thie effect, we repeat our analysis using 
narrower mass bins with $M_2=1.2M_1$. The results are 
plotted as the solid and dashed lines in Figure \ref{fig:fig2}. 
The results are almost indistinguishable from those obtained
with $M_2=2M_1$, demonstrating that the mass bins we used 
is sufficiently small.

\section{Summary and Discussion}

  Our major findings in this paper can be summarized as follows:
\begin{description}
 \item[(i)] The younger halo population is significantly less clustered
	    than the old one for $M<10M_\ast$;
 \item[(ii)] For $M>10M_\ast$, the age dependence is reversed, and 
the bias factor of the youngest 20\% of the halos with a given mass
is approximately 10\% {\it higher} than that of 
the oldest 20\% of the same mass; 
 \item[(iii)] When the halos are divided into subsamples according to
concentration parameter, the halo bias is larger
	    (smaller) for halos with higher concentrations for
	    $M<M_\ast$ ($M>M_\ast$).
\end{description}
The first result is a confirmation of the result of 
\citet{gao05} with the use of a much larger data set and a 
different clustering measure (our cross power spectrum versus
their auto-correlation function). Our use of a large set 
of simulations in large boxes enables us to quantify, 
for the first time, the age dependence of halo bias for 
$M >10 M_\ast$. The weak age dependence for massive halos 
relative to low-mass halos ones may reflect the fact 
that these halos have a narrower distribution in formation 
time \citep[see Fig.1 of][]{KS1996b}, but the opposite 
trend, although weak, is not easy to explain. The third result
is qualitatively in agreement with \citet{Wechsler06}.
However, their results are based on halos identified at 
different times, while ours are based on halos 
identified at the same tme as it should be. Our results 
also show that the concentration dependence is weaker than 
the age dependence for $M<M_{\ast}$ but stronger 
for $M\sim 50 M_{\ast}$. There is also indication that 
both the age and concentration dependence becomes weaker 
at the very massive end. 

The difference in the age and concentration dependence 
implies that caution must be taken in comparing model predictions 
with observational results 
\citep[e.g.][]{yang06}, since it is unclear whether 
the formation epoch, as defined by $z_f$, or concentration,
or even some other properties of a halo is more important in 
determining the properties of the galaxies that form in it.  

The dependence of halo bias on formation epoch and concentration 
implies that the simple excursion set theory of halo
formation \citep{ps74,bower91,bond91} is not accurate. 
As discussed in \citet{wang06}, the problem with the simple 
excursion set theory is that spherical collapse model, 
which neglects large-scale tidal field, over-predicts the collapse 
of small halos ($\ll M_{\ast}$) in high density environments. 
They argue, however, that this dynamical effect should be smaller 
for large halos ($M \gg M_{\ast}$). It is possible 
that the large-scale tidal field also plays a role in the 
formation of massive halos. But instead of truncating 
mass accretion, as is for the case of low-mass halos, the 
large-scale tidal field may delay the accretion, 
enhancing the accretion at later times. This possibility 
should be checked further by examining in detail the accretion 
histories and environments of massive halos in simulations.

  The extended PS theory \citep{LC1993,LC1994,sasaki94,KS1996a}
has been implemented in a variety of cosmological studies
with clusters of galaxies \citep{KS1996b,KS1997,taruya00,Hamana01}.
The present results for massive halos point to a 
limitation of the theory in such applications, 
and indeed should prompt the exploration of an improved
theoretical framework in the future.

\acknowledgments 
We thank Liang Gao for providing the data plotted in Figure 2.
YPJ is supported by NSFC(10373012, 10533030) and by Shanghai Key
Projects in Basic Research (04jc14079, 05xd14019).  The research of YS is
partly supported by Grant-in-Aid for Scientific Research of Japan
Society for Promotion of Science (No.14102004, 16340053).
HJM would like to acknowledge the support of NSF AST-0607535.
The visit of HJM to Shanghai Observatory was supported by the 
CAS scholar program.

\clearpage

\begin{deluxetable}{cccccc}
\tablewidth{0pt}
\tablecolumns{4}
\tablecaption{Simulation parameters}
\tablehead{\colhead{boxsize($\mpchi$)} 
& \colhead {particles} & \colhead {realizations} 
& \colhead {$m_{\rm particle}$}}
\startdata
300 & $1024^3$ & 1 & $1.8\times 10^{9}h^{-1}M_\odot$ \\
600 & $1024^3$ & 1 & $1.5\times 10^{10}h^{-1}M_\odot$\\
1200 & $1024^3$ & 1 & $1.2\times 10^{11}h^{-1}M_\odot$\\
1800 & $1024^3$ & 4 & $4.0\times 10^{11}h^{-1}M_\odot$
\enddata
\label{tab:simu}
\end{deluxetable}

\clearpage

\clearpage

\end{document}